\documentclass[aps,pre,twocolumn,showpacs,amsmath,amssymb,floatfix]{revtex4-1}
\usepackage[utf8x]{inputenc} 
\usepackage{mathtools}
\usepackage{graphicx}
\usepackage{multirow}
\usepackage{rotating}

\begin{document}

\title{Static critical behavior of the $q-$states Potts model: High-resolution
entropic study}

\author{A. A. Caparica}
\email{caparica@ufg.br}
\author{Salviano A. Le\~ao}
\affiliation{Instituto de F\'{\i}sica, Universidade Federal de Goi\'as,
C.P. 131 CEP 74001-970, Goi\^{a}nia, Goi\'{a}s, Brazil}
\author{Claudio J. DaSilva}
\affiliation{Instituto Federal de Educa\c{c}\~ao, Ci\^encia e Tecnologia de
Goi\'as, CEP 74130-012, Goi\^ania, Goi\'{a}s, Brazil}

\begin{abstract}
Here we report a precise computer simulation study of the static critical
properties of the two-dimensional $q$-states Potts model using very accurate
data obtained from a modified Wang-Landau (WL) scheme proposed by Caparica and
Cunha-Netto [Phys. Rev. E {\bf 85}, 046702 (2012)]. This algorithm is an
extension of the conventional WL sampling, but the authors changed the criterion
to update the density of states during the random walk and established a new
procedure to windup the simulation run. These few changes have allowed a more
precise microcanonical averaging which is essential to a reliable finite-size
scaling analysis. In this work we used this new technique to determine the
static critical exponents $\beta$, $\gamma$, and $\nu$, in an unambiguous
fashion. The static critical exponents were determined as $\beta=0.10811(77)$,
$\gamma=1.4459(31)$, and $\nu=0.8197(17)$, for the $q=3$ case, and
$\beta=0.0877(37)$, $\gamma=1.3161(69)$, and $\nu=0.7076(10)$, for the $q=4$
Potts model. A comparison of the present results with conjectured values and
with those obtained from other well established approaches strengthens this new
way of performing WL simulations.
\end{abstract}

\maketitle

\section{Introduction}

Monte Carlo (MC) simulations are ubiquitous in the field of statistical
mechanics, especially for the study of phase transitions and critical
phenomena \cite{Binder2000,krauth2006}. Since the historical work of Metropolis
\textit{et al} \cite{Metropolis1953}, the most outstanding task in this context
is the pursuit of new and more efficient algorithms to overcome long time scale
problems. Since there are few problems in the field of interacting systems for
which one can find an exact solution, MC simulations became an indispensable
tool. This is due to the massively increasing in computational power and further
due to the development of more efficient algorithms. More recently, such
development focused on the extended ensemble method, where one uses an ensemble
different from the ordinary canonical with a fixed temperature, as in the
original Metropolis algorithm. To name a fill examples we have the
multicanonical method \cite{BergNeuhaus1992}, and the exchange Monte Carlo method
(parallel tempering) \cite{Nemoto1996}. Particularly, during the last two
decades, a multicanonical MC algorithm known as Wang-Landau
sampling \cite{WangLandau2001}, has been at the forefront of
interest \cite{Dayaletal2004} and has proven to be a very powerful numerical
procedure for the study of phase transitions and critical
phenomena \cite{Landau2004,Nogawa2001,blume}.

The original idea of the WL algorithm is to measure an {\it a priori} unknown
density of states of a given system iteratively by performing a random walk in
energy space and sampling configurations with probability proportional to the
reciprocal of the density of states, resulting in a ``flat" histogram. Despite
being a well-established numerical procedure, it is clear that some improvements
on the algorithm are indeed necessary to overcome some limitations during the
simulation run. The method itself was subject to several studies and various
improvements to it have been
proposed \cite{kawashima,ZhouBhatt2005,belardinelli}. By its turn, the MC
algorithm used in this work is an extension of the conventional WL where some
few changes produce more reliable and precise results.

Considering the aforementioned comments, the purpose of this paper is twofold.
First, to present a numerically simple and accurate procedure to halting a
regular WL simulation run. This is accomplished with a method proposed in
Refs. \cite{caparica,caparica3}. Second, to apply this technique to the square
two-dimensional $q$-state Potts model and compute the static critical exponents
for $q=3$ and $4$ states, showing that this method is also a helpful tool to
address the achievement of critical exponents, a possibility barely explored in
the literature, the exception being the important works of Malakis {\it et al}
 \cite{malakis1,malakis2,malakis3,malakis4}. In the following we will make use of
a combination between finite-size scaling theory and cumulant methods to locate
and evaluate the extrema of various thermodynamic quantities and estimate the
static critical exponents.

The outline of this paper is as follows: In section II we define the model . In
section III we define the simulation procedure. In section IV we describe the
finite-size scaling analysis. The results are discussed in section V. Section VI
is devoted to the summary and concluding remarks.

\section{$q$-states Potts Model}

The Potts model, proposed by Potts in the early 1950's, has stood at the frontier
of research in statistical mechanics since its formulation. It is an extension
of the two states Ising model to $q>2$ states. In this model, to each  lattice site
is attached a spin variable $\sigma_i$ (defined on each site $i$) which
takes on integer values $1, \ldots, q$. Adjacent sites have an attractive
interaction energy $-J$ whenever they are equal or $0$ otherwise. The
Hamiltonian of the q-states  ferromagnetic model ($J>0$) can be written as
 \cite{barkema}
\begin{equation}
  {\cal H} = -J \sum_{<i,j>} \delta_{\sigma_i  \sigma_j},
\end{equation}
where $\delta$ is the Kronecker $\delta-$symbol, and the sum runs over all
nearest neighbors of $\sigma_i$. In the low temperature regime the system is
ordered, becoming disordered as $T$ increases. In 2D, for $q \leq 4$ the phase
transition is of second-order and discontinuous if $q \geq 5$. A proper order
parameter $\phi$ is
\begin{equation}
  \phi = \frac{q (N_{max} / N)-1}{q-1},
  \label{order-potts}
\end{equation}
where $N_{max}$ is the ``volume" occupied by the spins of the state $q$ of
largest population and $N = L^2$ \cite{Wu1982}.

\section{Entropic Simulations}

The Wang-Landau method \cite{WangLandau2001} is based on the fact that if one
performs a random walk in energy space with a probability proportional to the
reciprocal of the density of states, a flat histogram is generated for the
energy distribution. Since the density of states produces huge numbers, instead
of estimating $g(E)$, the simulation is performed for $S(E)\equiv\ln g(E)$. At
the beginning we set $S(E)=0$ for all energy levels. The random walk in the
energy space runs through all energy levels from $E_{min}$ to $E_{max}$ with a
probability $p(E\rightarrow E')=\min(\exp{[S(E)-S(E')]},1)$, where $E$ and $E'$
are the energies of the current and the new possible configurations,
respectively. Whenever a configuration is accepted we update $H(E')\rightarrow
H(E')+1$ and $S(E')\rightarrow S(E')+F_{i}$, where $F_{i}=\ln f_{i}$,
$f_{0}\equiv e=2.71828...$ and $f_{i+1}=\sqrt{f_{i}}$ ($f_{i}$ is the so-called
modification factor). The flatness of the histogram is checked after a certain number of
Monte Carlo steps (MCS) and usually the histogram is considered flat if
$H(E)>0.8\langle H \rangle$, for all energies, where $\langle H \rangle$ is an
average over energies. If the flatness condition is fulfilled we update the
modification factor to a finer one and reset the histogram $H(E)=0$.

Recent works \cite{caparica,caparica3,caparica1,caparica2} have demonstrated
that (a) instead of updating the density of states after every move, one ought
to update it after each Monte Carlo sweep \cite{mcs}(this providence avoids
taking into account highly correlated configurations when constructing the density
of states); (b) WL sampling should be carried out only up to $\ln f=\ln f_{final}$ defined
by the canonical averages during the simulations (this saves CPU time, discarding
unnecessary long simulations); and (c) the microcanonical averages should not be
accumulated before $\ln f \leq ln f_{micro}$ defined by a previous study of the
microcanonical averaging during the simulation (the ruled out WL levels in these averages correspond
to a microcanonical termalization, since the initial configurations do not match those
of maximum entropy). The adoption of these easily implementable
changes leads to more accurate results and saves computational time. They
investigated the behavior of the maxima of the specific heat
\begin{equation}\label{cv}
 C(T)=\langle(E-\langle E\rangle)^2\rangle/T^2
\end{equation}
and the magnetic susceptibility
\begin{equation}\label{ki}
 \chi(T)=L^2\langle(m-\langle m\rangle)^2\rangle/T,
\end{equation}
where $E$ is the energy of a given configuration and $m$ is the corresponding
magnetization per spin, during the WL sampling for the Ising model on a square lattice.
They observed that a considerable part of the conventional Wang-Landau
simulation is not very useful because the error saturates. They demonstrated in
detail that in general no single simulation run converges to the true value, but
to a particular value of a Gaussian distribution of results around the correct
value. The saturation of the error coincides with the convergence to this value.
Continuing the simulations beyond this limit leads to irrelevant variations in
the canonical averages of all thermodynamic variables.

Zhou and Bhatt \cite{ZhouBhatt2005} demonstrated that when $f$ is
close to $1$ the relative error $\delta g/g=\delta\ln g$ scales as $\sqrt{\ln f}$.
Conversely, in Ref. \cite{caparica3} it was shown that this convergence indeed holds, but
the final result falls in a Gaussian distribution around the true value. In this work
it is also noteworthy that the convergence described in Ref. \cite{dickman} for the
$1/t$ entropic sampling, where the authors argue that the logarithm of the density
of states converges as $1/\sqrt{t}$, is not reflected in the canonical averages,
since for long simulations different runs do not converge to a unique value, moreover
the results take on an erratic behavior.

Ref. \cite{caparica3} also proposes a criterion for halting the simulations.
Applying WL sampling to a given model,
beginning from $f_{5}$, we calculate the temperature of the peak of the specific
heat defined in Eq. \eqref{cv} using the current $g(E)$ and from this time forth
this mean value is updated whenever the histogram is checked for flatness. When
the histogram is considered flat, we save the value of the temperature $T_c(0)$
of the peak of the specific heat. We then update the modification factor
$f_{i+1}=\sqrt{f_{i}}$ and reset the histogram $H(E)=0$. During the simulations
with this new modification factor we continue calculating the temperature of the
peak of the specific heat $T_c(t)$ whenever we check the histogram for flatness
and we also calculate the checking parameter

\begin{equation}\label{eps}
 \varepsilon=|T_c(t)-T_c(0)|.
\end{equation}

If the number of MCS before verifying the histogram for flatness is chosen not
too large, say 10,000, then during the simulations with the same modification
factor the checking parameter $\varepsilon$ is calculated many times. If
$\varepsilon$ remains less than $10^{-4}$ until the histogram meets the flatness
criterion for this WL level, then we save the density of states and the
microcanonical averages and stop the simulations. When one adopts this criterion
for halting the simulations, different runs stop at different final modification
factors.

Having at hand the density of states, one can calculate the canonical average of
any thermodynamic variable $X$ as
\begin{equation}\label{mean}
\langle X\rangle_T=\dfrac{\sum_E \langle X\rangle_E g(E) e^{-\beta E}}{\sum_E g(E) e^{-\beta E}} ,
\end{equation}
where $\langle X\rangle_E$ is the microcanonical average accumulated during the
simulations and $\beta=1/k_BT$, where $T$ is the absolute temperature measured
in units of $J/k_B$ and $k_B$ is the Boltzman's constant.

In Ref. \cite{caparica3} it was also observed that two independent similar
finite-size scaling procedures can lead to very different results for the
critical temperature and exponents, which often do not agree within the error
bars. The way to overcome this difficulty is to carry out 10 independent sets of
finite-size scaling simulations. In the present work, for each of these sets
and for each Potts model ($q=3$ and $q=4$), we performed simulations for $L=
32,36,40,44,48,52,56,64,72,$ and $80$ with $n=24,24,20,20,20,16,16,16,12,$ and $12$
independent runs for each size, respectively. The final resulting values for the
critical exponents were obtained as an average over all sets.

\section{Finite-size scaling}

According to finite-size scaling theory \cite{fisher1,fisher2,barber} from the
definition of the free energy one can obtain the zero field scaling expressions
for the magnetization, susceptibility, and specific heat, respectively, by

\begin{equation}\label{exp1}
 m\approx L^{-\beta/\nu}\mathcal{M}(tL^{1/\nu}),
\end{equation}

\begin{equation}\label{exp2}
\chi \approx L^{\gamma/\nu}\mathcal{X}(tL^{1/\nu}).
\end{equation}

\begin{equation}\label{exp3}
c \approx c_\infty + L^{\alpha/\nu}\mathcal{C}(tL^{1/\nu}),
\end{equation}
where $t=(T_c-T)/T_c$ is the reduced temperature, and $\alpha$, $\beta$, and
$\gamma$ are static critical exponents which should satisfy the scaling
relation \cite{privman}

\begin{equation}\label{scaling_relation}
 2-\alpha=d\nu=2\beta+\gamma.
\end{equation}

The critical temperature for the Potts model (for $q\geq 4$) is exactly known as

\begin{equation}\label{tcpotts}
\frac{k_BT_c}{J}=\frac{1}{\ln(1+\sqrt{q})}
\end{equation}
and it is expected that this expression is also exact for $q=3$, although a
rigorous proof of this assumption is still lacking \cite{Wu1982}.

Following Refs. \cite{chen1993,double} we can define a set of thermodynamic
quantities related to logarithmic derivatives of the magnetization:

\begin{spreadlines}{0.9em}
\begin{align}
V_1 & \equiv 4[m^3]-3[m^4],  \label{v1} \\
V_2 & \equiv 2[m^2]-[m^4],   \label{v2} \\
V_3 & \equiv 3[m^2]-2[m^3],  \label{v3} \\
V_4 & \equiv (4[m]-[m^4])/3, \label{v4} \\
V_5 & \equiv (3[m]-[m^3])/2, \label{v5} \\
V_6 & \equiv 2[m]-[m^2],     \label{v6}
\end{align}
\end{spreadlines}
where

\begin{equation}\label{mn}
[m^n] \equiv \ln \frac{\partial \langle m^n \rangle}{\partial T}.
\end{equation}

Using Eq. \eqref{exp1} it is easy to show that

\begin{equation}\label{mn}
V_j\approx \frac{1}{\nu}\ln L+\mathcal{V}_j(tL^{1/\nu})
\end{equation}
for $j=1,2,...,6.$ Since the critical temperature $T_c$ is known for both
models, at the critical temperature $t=0$ and the $\mathcal{V}_j$ are constants
independent of the system size and we can estimate $1/\nu$ by the slopes of
$V_j$ calculated at $T_c$. And then, with the exponent $\nu$ at hand, we can
estimate the exponents $\beta$ and $\gamma$ by the slopes of the log-log plots
of Eqs. \eqref{exp1} and \eqref{exp2} calculated at the critical temperature
$T_c$.

\section{Results}

\begin{figure}[!h]
\begin{center}
 \includegraphics[width=.90\linewidth]{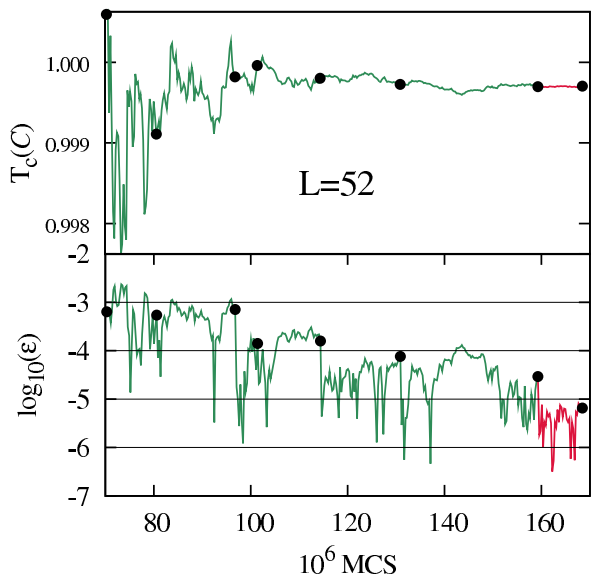}
\end{center}
\vspace{-.5cm}
\caption{(color online). Upper panel: Evolution of the temperature of the
maximum of the specific heat during the WL sampling, beginning from $f_{9}$ for a single
run. The dots show where the modification factor was updated. Lower panel:
Evolution of the logarithm of the checking parameter $\varepsilon$ during the
same simulation.} \label{checking}
\end{figure}

In all simulations we carried out, the microcanonical averages were accumulated
beginning from $f_7$, we adopted the MCS for updating the density of states and
the jobs were halted using the checking parameter $\varepsilon$. In Fig.
\ref{checking} we show the evolution of the temperature of the maximum of the
specific heat during the WL sampling beginning from $f_9$ for a single run with $L=52$
and the evolution of log$_{10}(\varepsilon)$ during the same simulation. One can
see that at the last WL level the logarithm of $\varepsilon$ remains less than
-4 indicating that the simulation can be stopped at the end of $f_{15}$.

According to Eq. \eqref{tcpotts} the critical adimensional temperature for the
$q=3$ Potts model is given by

\begin{equation}\label{tc}
\frac{k_BT_c}{J}=\frac{1}{\ln(1+\sqrt{3})}=0.994972861...
\end{equation}

\begin{figure}[!ht]
\begin{center}
 \includegraphics[width=.70\linewidth, angle=-90]{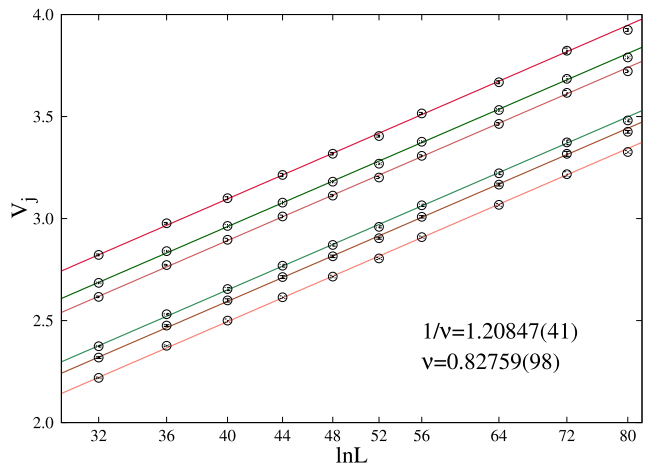}
\end{center}
\vspace{-.5cm}
\caption{(color online) Size dependence of $V_j$ at the critical temperature. The slopes yield $1/\nu$.}
\label{vj}
\end{figure}

Evaluating the thermodynamic quantities Eqs. \eqref{v1}-\eqref{v6} at this
temperature and taking into account Eq. \eqref{mn}, we are able to determine
$\frac{1}{\nu}$ by the slopes of the straight lines that we obtain with respect
to $\ln L$.  For each of these six slopes we calculate $\nu=1/(\frac{1}{\nu})$
with $\varDelta\nu=\varDelta(\frac{1}{\nu})/(\frac{1}{\nu})^2$ and take an
average with unequal uncertainties \cite{wong} over them. In Fig. \ref{vj} we
present this set of lines. From the linear fits to these points we estimate that
$\frac{1}{\nu}=1.20847(41)$, yielding ${\nu}=0.82759(98)$. Nevertheless these
values represent the result of only one of the $10$ sets of finite-size scaling
simulations which were carried out. Initially we run over all sets calculating
$\nu$ in order to determine this exponent to the best precision.

At this point we take a moment to discuss which procedure should be adopted
to calculate the mean value of these $10$ results, a single averaging or an average with
unequal uncertainties. In order to investigate the behavior of the data under
these two procedures, we grouped the five first sets in a large one and the last
five in another large set. Taking the averages with unequal uncertainties
we obtained $\nu=0.82272(26)$ and $0.81658(38)$, while if we take just single
averages neglecting the error bars, we obtain $\nu=0.8230(19)$ and $0.8165(22)$, in
each of these two large sets. One can see that the former procedure leads to
unrealistic error bars, whereas the later yields results that intersect within
$\pm2\sigma$ errors. We therefore adopt the single averaging here and in all the
further calculations. In Table \ref{table1} the fourth column displays the values
obtained in each set and the final result in the last line: ${\nu}=0.8197(17)$.

\begin{figure}[!ht]
\begin{center}
 \includegraphics[width=.70\linewidth, angle=-90]{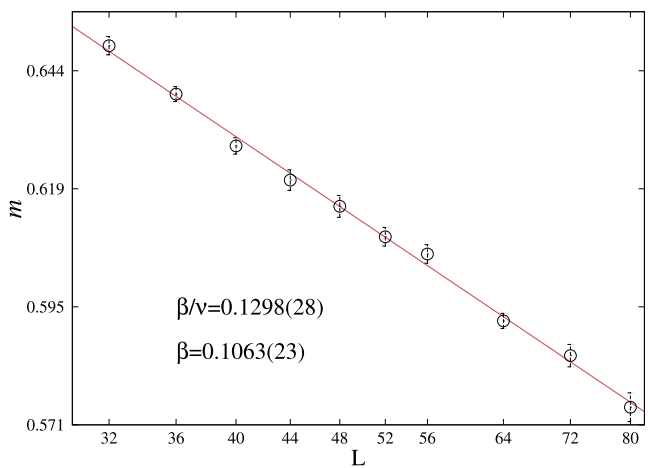}
\end{center}
\vspace{-.5cm}
\caption{(color online) Log-log plot of size dependence of the magnetization at $T_c=0.994972861$.}
\label{beta}
\end{figure}

\begin{figure}[!ht]
\begin{center}
 \includegraphics[width=.70\linewidth, angle=-90]{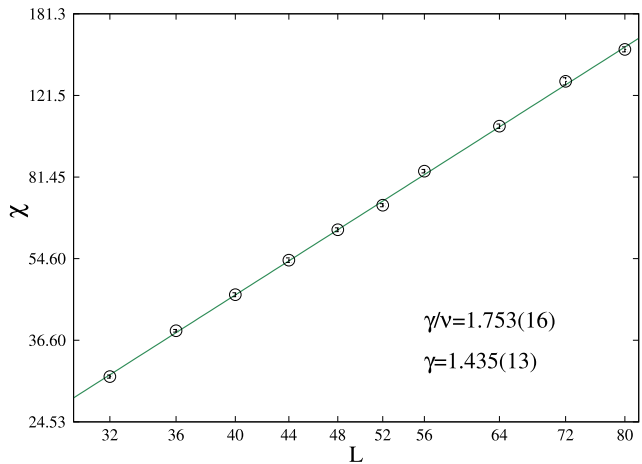}
\end{center}
\vspace{-.5cm}
\caption{(color online) Log-log plot of size dependence of the susceptibility at $T_c=0.994972861$.}
\label{gama}
\end{figure}

Next, with the critical exponent $\nu$ accurately determined, we can use
Eqs.\eqref{exp1}-\eqref{exp2} to evaluate the exponents $\frac{\beta}{\nu}$ and
$\frac{\gamma}{\nu}$ by the slopes of the log-log plots. In Fig. \ref{beta} and
Fig. \ref{gama} we show this finite-size scaling behavior for each exponent,
obtaining $\frac{\beta}{\nu}=0.1298(28)$ and $\frac{\gamma}{\nu}=1.753(16)$,
respectively. We then calculate $\beta=\nu\frac{\beta}{\nu}$ with
$\varDelta\beta= \frac{\beta}{\nu}\varDelta\nu+\nu\varDelta\frac{\beta}{\nu}$,
and similarly for $\gamma$ and $\varDelta\gamma$, obtaining $\beta=0.1063(23)$,
and $\gamma=1.435(13)$. Again, these values were obtained at the first folder.
In Table \ref{table1} we show the results for all sets and the best estimates in
the last line, yielding $\beta=0.10811(77)$, and $\gamma=1.4459(31)$. Finally
using the scaling relation given by Eq. \eqref{scaling_relation} we determined
the exponent $\alpha=2-2\beta-\gamma$ with
$\varDelta\alpha=2\varDelta\beta+\varDelta\gamma$. These results are also
displayed in Table \ref{table1} giving $\alpha=0.3379(28)$.

\begin{table*}[!ht]
\begin{ruledtabular}
  \begin{tabular}{c c c c c c l c l}
    $\alpha$ &  $\beta$ & $\gamma$ & $\nu$ & \phantom{x} &$\alpha$ &
    \multicolumn{1}{c}{$\beta$} & $\gamma$ & \multicolumn{1}{c}{$\nu$}  \\ \cline{1-4} \cline{6-9}
    \multicolumn{4}{c}{$q=3$ Potts model} &  &\multicolumn{4}{c}{$q=4$ Potts model} \\ \cline{1-4} \cline{6-9}
    0.352(17) & 0.1063(23) & 1.435(13)  & 0.82759(98) & & 0.550(26) & 0.0836(71) & 1.283(12) & 0.7123(13)  \\
    0.351(12) & 0.1063(31) & 1.4367(62) & 0.82364(52) & & 0.541(18) & 0.0951(35) & 1.269(11) & 0.7045(13)  \\
    0.340(14) & 0.1120(32) & 1.4361(83) & 0.82068(56) & & 0.547(35) & 0.0855(74) & 1.282(20) & 0.70907(91) \\
    0.334(14) & 0.1044(37) & 1.4569(69) & 0.81688(60) & & 0.539(24) & 0.0950(43) & 1.271(16) & 0.7085(12)  \\
    0.350(16) & 0.1093(39) & 1.4315(91) & 0.82601(47) & & 0.504(33) & 0.0907(74) & 1.315(19) & 0.70185(77) \\
    0.341(16) & 0.1099(23) & 1.440(12)  & 0.81797(88) & & 0.519(27) & 0.0895(50) & 1.302(17) & 0.7041(12)  \\
    0.337(12) & 0.1099(30) & 1.4428(66) & 0.82410(79) & & 0.538(24) & 0.0946(66) & 1.273(11) & 0.7085(13)  \\
    0.336(12) & 0.1053(33) & 1.4535(62) & 0.81632(63) & & 0.528(15) & 0.0898(28) & 1.292(10) & 0.70671(83) \\
    0.326(13) & 0.1094(27) & 1.4552(85) & 0.81162(52) & & 0.521(23) & 0.0958(50) & 1.287(14) & 0.7059(14)  \\
    0.329(10) & 0.1071(27) & 1.4563(50) & 0.81249(45) & & 0.550(46) & 0.056(12)  & 1.337(24) & 0.7074(12)  \\ \cline{1-4} \cline{6-9}
    0.3379(28)& 0.10811(77)& 1.4459(31) & 0.8197(17) & & 0.5084(48)& 0.0877(37) &1.3161(69) & 0.7076(10) \\

\end{tabular}
\end{ruledtabular}
  \caption{$q=3$ and $q=4$ Potts models: Ten finite-size scaling results for the
exponents $\alpha$, $\beta$, $\gamma$, and $\nu$. The last line shows the
average values over all the runs.}
  \label{table1}
\end{table*}

\begin{table*}[!hbt]
\begin{ruledtabular}
\begin{tabular}{l c c c c}
Method                                  & $\alpha$     & $\beta$        & $\gamma$       & $\nu$          \\
\hline
\multicolumn{5}{c}{$q=3$ Potts model} \\ \hline
Conjectured value \cite{Wu1982}         & $\frac{1}{3}$& $\frac{1}{9}$  & $\frac{13}{9}$ & $\frac{5}{6}$  \\
Kadanoff variational RG \cite{dasgupta} & $0.326$      & $0.107$        & $1.460$        & $0.837$        \\
Monte Carlo RG \cite{rebbi}             & $0.352$      & $0.101$        & $1.445$        & $0.824$        \\
This work                               & $0.3379(28)$ & $0.10811(77)$  & $1.4459(31)$ & $0.8197(17)$   \\  \hline
\multicolumn{5}{c}{$q=4$ Potts model} \\ \hline
Conjectured value \cite{Wu1982}         & $\frac{2}{3}$& $\frac{1}{12}$ & $\frac{7}{6}$  & $\frac{2}{3}$  \\
Kadanoff variational RG \cite{dasgupta} & $0.488$      & $0.091$        & $1.330$        & $0.756$        \\
Duality invariant RG \cite{hu}          & $0.4870$     & $-$            & $-$            & $0.7565$       \\
This work                               & $0.5084(48)$ & $0.0877(37)$   & $1.3161(69)$   & $0.7076(10)$   \\
\end{tabular}
\end{ruledtabular}
\caption{Estimates of $\alpha$, $\beta$, $\gamma$, and $\nu$ compared to results obtained with other techniques and conjectured
values.}
\label{table2}
\end{table*}

For the $q=4$ Potts model the critical adimensional temperature is given by

\begin{equation}\label{tc}
\frac{k_BT_c}{J}=\frac{1}{\ln(1+\sqrt{4})}=0.910239226...
\end{equation}

All the plots and finite-size scaling procedures are completely analogous to
those we described above for the $q=3$ case. In Table \ref{table1} we display
the results for the 10 folders and our final estimates yielding
$\alpha=0.5084(48)$, $\beta=0.0877(37)$, $\gamma=1.3161(69)$, and
$\nu=0.7076(10)$.

Such large repetitious handling of data for obtaining all these canonical
averages and  finite-size scaling extrapolations were possible only by using
shell scripting \cite{AdvBashScr,BGB2008,Robbins2005,Neves2008,Jargas2008-Shell}.
This is an exceptional tool for those who work with simulations.

As a final discussion, we compare in Table \ref{table2} our final estimates 
of the critical exponents to other well-established values. It is possible 
to see a good agreement for the $q=3$ case, especially between those obtained 
by numerical means. For the $q=4$ case, our results are below those of Refs. 
\cite{dasgupta,hu}. Notwithstanding our results for $\beta$, $\gamma$ and $\nu$
are closer to the conjectured ones when compared to these approaches. 
This is a clear indication that our procedure of carefully handling very accurate 
data obtained by an entropic sampling simulation is a powerful and reliable technique.

\section{Conclusions}

In this work we explored the static critical behavior of $q=3$ and $q=4$ Potts
models within a high-precision and refined Wang-Landau procedure. All results are
in very good agreement with those obtained from other well established
approaches. The most striking conclusion from our analysis, in our opinion, is
that it is possible to obtain reliable and very precise calculations of critical
exponents from WL sampling provided that the appropriate implementations
adopted in this work are made. Most important, the implementation of the
present method remains as simple as the original idea of the WL algorithm. A further critical
test of our algorithm would be provided by an analysis of the critical behavior
of multi-parametric spin systems, which is a hard task for any conventional WL
approach.

\section{Acknowledgment}

This work was supported by FUNAPE-UFG. We acknowledge the computer resources provided by LCC-UFG.


\begin{thebibliography}{99}

\bibitem{Binder2000}{D.P. Landau, K. Binder, A Guide to Monte Carlo Simulations in Statistical Physics, Cambridge University Press, New York, USA, 2000.}

\bibitem{krauth2006}{W. Krauth, Statistical Mechanics: Algorithms and Computations, Oxford University Press, USA, 2006.}

\bibitem{Metropolis1953}{N. Metropolis, A. Rosenbluth, M. Rosenbluth, A. Teller and E. Teller, J. Chem. Phys. {\bf 21}, 1087 (1953).}

\bibitem{BergNeuhaus1992}{B. A. Berg and T. Neuhaus, Phys. Rev. Lett. {\bf 68}, 9 (1992).}

\bibitem{Nemoto1996}{K. Hukushima and K. Nemoto, J. Phys. Soc. Jpn. {\bf 65}, 1604 (1996).}

\bibitem{WangLandau2001}{F. Wang and D. P. Landau, Phys. Rev. Lett. {\bf 86}, 2050 (2001); Phys. Rev. E {\bf 64}, 056101 (2001).}

\bibitem{Dayaletal2004}{P. Dayal, S. Trebst, S. Wessel, D. Wurtz, M. Troyer, S. Sabhapandit, and S. N. Coppersmith, Phys. Rev. Lett. {\bf 92}, 097201 (2004).}

\bibitem{Landau2004}{D. P. Landau, S.-Ho Tsai, and M. Exler, Am. J. Phys. {\bf 72}, 1294 (2004).}

\bibitem{Nogawa2001}{T. Nogawa, N. Ito, and H. Watanabe, Phys. Rev. E {\bf 84}, 061107 (2011).}Nogawa,T./Ito,N.//

\bibitem{blume} C. J. Silva, A. A. Caparica and J. A. Plascak, Phys. Rev. E \textbf{73}, 036702 (2006).

\bibitem{kawashima} C. Yamaguchi and N. Kawashima, Phys. Rev. E \textbf{65}, 056710 (2002).

\bibitem{ZhouBhatt2005}{C. Zhou and R. N. Bhatt, Phys. Rev. E {\bf 72}, 025701 (2005).}

\bibitem{belardinelli} R. E. Belardinelli and V. D. Pereyra, J. Chem. Phys. \textbf{127}, 184105 (2007).

\bibitem{barkema} M.E.J. Newman and G.T. Barkema, \textit{Monte Carlo Methods in Statistical Physics}, Claredon Press, Oxford,
p. 120 (2001).

\bibitem{Wu1982}{F.Y Wu, The Potts model, Rev. Mod. Phys. {\bf 54}, 235 (1982).}

\bibitem{dickman} Belardinelli, R.E., Pereyra, V.D., Dickman, R., and Lourenço, B.J., J. Stat. Mech., \textbf{2014}, P07007 (2014).

\bibitem{caparica} A.A. Caparica and A.G. Cunha-Netto, Phys. Rev. E \textbf{85}, 046702 (2012).

\bibitem{caparica3} A.A. Caparica, Phys. Rev. E \textbf{89}, 043301 (2014).

\bibitem{caparica1}L.S. Ferreira and A.A. Caparica, Int. J. Mod. Phys. C \textbf{23}, 1240012 (2012).

\bibitem{caparica2}L.S. Ferreira, A.A. Caparica, M. A. Neto, and M. D. Galiceanu, J. Stat. Mech., \textbf{2012}, P10028 (2012).

\bibitem{malakis1} A. Malakis, A. Peratzakis, and N. G. Fytas, Phys. Rev. E {\bf 70}, 066128 (2004).

\bibitem{malakis2} A. Malakis, S.S. Martinos, I.A. Hadjiagapiou, N. G. Fytas, and P. Kalozoumis, Phys. Rev. E {\bf 72}, 066120 (2005).

\bibitem{malakis3} A. Malakis, A. N. Berker, I. A. Hadjiagapiou, and N.G. Fytas, Phys. Rev. E {\bf 79}, 011125 (2009).

\bibitem{malakis4} A. Malakis, A. N. Berker, I. A. Hadjiagapiou, N. G. Fytas, and T. Papakonstantinou, Phys. Rev. E {\bf 81}, 041113 (2010).

\bibitem{mcs} A Monte Carlo sweep consists of $L^2$ spin-flip trials in the 2D Ising model or $N$
monomer moves in the homopolymer.

\bibitem{fisher1} M.E. Fisher, in \textit{Critical Phenomena}, edited by M. S. Green (Academic, New York, 1971).

\bibitem{fisher2} M.E. Fisher and M.N. Barber, Phys. Rev. Lett. \textbf{28}, 1516 (1972).

\bibitem{barber} \textit{Phase Transitions and Critical Phenomena}, edited by C. Domb and J. L. Lebowitz
(Academic, New York, 1974), Vol. 8.

\bibitem{privman} V. Privman, P.C. Hohenberg, and A. Aharony, in \textit{Phase Transitions and Critical Phenomena},
eduted by C. Domb and J. L. Lebowitz (Academic, New York, 1991), Vol. 14, p. 1.

\bibitem{chen1993} K. Chen,A.M. Ferrenberg, and D.P. Landau, Phys. Rev. B\textbf{48}, 3249 (1993).

\bibitem{double}A.A. Caparica, A. Bunker, and D.P. Landau, Phys. Rev. B \textbf{62}, 9458 (2000); There is a misprinting
in Eq.(3) in this paper, which should be $V_5\equiv (3[m]-[m^3])/2$.

\bibitem{wong}S.S.M. Wong, \textit{Computational Methods in Physics and Engineering}, 2\textit{nd} edition, World Scientific Publishing Co. Pte. Ltd. (1997).

\bibitem{dasgupta} C. Dasgupta, Phys. Rev. B {\bf 15}, 3460 (1977).

\bibitem{rebbi} C. Rebbi, and R. H. Swendsen, Phys. Rev. B {\bf 21}, 4094 (1980).

\bibitem{hu} B. Hu, J. Phys. A {\bf 13}, L321 (1980).

\bibitem{AdvBashScr} M. Cooper,
  ``\href{http://www.tldp.org/LDP/abs/html/}{Advanced bash-scripting guide}'' (2012).
  Retrieved from {\url{http://www.tldp.org/LDP/abs/html/}}.

\bibitem {BGB2008}  M. Garrels
  ``\href{http://www.tldp.org/LDP/Bash-Beginners-Guide/html/index.html}{Bash guide for beginners}''
  (2008). Retrieved from {\url{http://www.tldp.org/LDP/Bash-Beginners-Guide/html/index.html}}.

\bibitem{Robbins2005} A. Robbins and N.H.F. Beebe,
 ``Classic Shell Scripting: Hidden Commands that Unlock the Power of Unix'',
  Oreilly Series, (O'Reilly 2005).

\bibitem{Neves2008} J.C. Neves, ``Programa{\c{c}}{\~a}o Shell Linux'',
$7^{\underline{a}}$ edi{\c{c}}{\~a}o, (Brasport, 2008).

\bibitem{Jargas2008-Shell} A.M. Jargas, ``Shell Script Profissional'' (Novatec, 2008).

\bibitem{nauenberg}M. Nauenberg, and D.J. Scalapino, Phys. Rev. Lett. \textbf{44}, 837 (1980).

\bibitem{salas1997}{J. Salas ,and A. D. Sokal, J. Stat. Phys. {\bf 88}, 567 (1997).}

\bibitem{fytas2012}{P. E. Theodorakis, and N. G. Fytas, Phys. Rev. E {\bf 86}, 011140 (2012).}

\end{thebibliography}
\end{document}